\documentstyle[12pt]{article}
\begin{document}
\setlength{\topmargin}{1.0cm}
 \newcommand\beq{\begin{equation}}
  \newcommand\eeq{\end{equation}}
 \newcommand\beqn{\begin{eqnarray}}
  \newcommand\eeqn{\end{eqnarray}}
 \newcommand{\doublespace} {
  \renewcommand{\baselinestretch} {1.6} \large\normalsize}
  \renewcommand{\thefootnote}{\fnsymbol{footnote}}

 \setlength{\baselineskip}{0.55cm}

\begin{center}

 {\huge{\bf Hadronization in Nuclear
 Environment\\
and\\

 \medskip

  Electroproduction of Leading
 Hadrons}\footnote{Invited talk given by B.~Kopeliovich
at the ELFE (Electron Lab. For Europe) Workshop,\\
Cambridge, 22-29 July, 1995}} \\

\vspace{1.5cm}

 {\large Boris~Kopeliovich}\footnote{On leave from
 Joint
Institute for Nuclear Research, Dubna, 141980
 Moscow Region, Russia.\\
E-mail: bzk@dxnhd1.mpi-hd.mpg.de}\\
 \medskip

 {\it Max-Planck
Institut f\"ur Kernphysik, Postfach 103980,
 \newline 69029 Heidelberg,
Germany}\\
 \bigskip

 {\large Jan Nemchik\footnote{On leave from
Institute of Experimental Physics SAV,
 Solovjevova 47, CS-04353 Kosice,
Slovakia}~
 and Enrico Predazzi}\\
 \medskip

 {\it Dipartimento di Fisica
Teorica,
 Universit\`a di Torino\\
 and INFN, Sezione di Torino, I-10125,
Torino, Italy}
 \\
\vspace{1.5cm}

 \large {\bf ABSTRACT}

\end{center}

Radiative energy loss of a highly virtual quark
originating from a deep-inelastic electron scattering
plays a crucial role in production of leading hadrons
off nuclei. The density of energy loss for gluon radiation
turns out to be time- and energy--dependent in inclusive
hadron production. Important phenomena involved
are Sudakov's suppression of no radiation of that part of
gluon spectrum, which is forbidden by energy conservation,
and color transparency, which suppresses the final state
interaction of the produced colorless wave packet.

We model the soft part of
hadronization, which usually is supposed to be due to
the color strings, using also gluon radiation
and nicely
reproduce the string parameters.

Our parameter-free calculations provide a good
agreement with available data on $z_h$-dependence
of the quark fragmentation function in vacuum,
as well as $\nu$-, $z_h$-
and $Q^2$-dependence of nuclear effects.
We come to the conclusion that the energy
range of ELFE - HERMES is most sensitive to the
underlying dynamics of hadronization provided
that nuclear targets are used.
\vspace{2.0cm}

\pagebreak

 \setlength{\topmargin}{-1.5cm}
\setlength{\textheight}{25.0cm}
 \setlength{\baselineskip}{0.75cm}

\section{Why nuclear target?}

 A quark originated from a hard reaction,
eventually converts into colorless hadrons due to
 confinement.  The Lorenz
time dilation stretches
 considerably the duration of this process, and the
hadrons carry poor information about the early
 stage of hadronization.  A
nuclear target provides
 a unique opportunity to look inside the process at
very short times after it starts.  The quark-gluon
 system produced in a hard
collision, interacts
 while passing through the nucleus.  Observation of
manifestations of that can bring forth precious
 information about the
structure and the space-time
 pattern of hadronization.

 The modification
of the quark fragmentation
 function by nuclear matter was considered for
high-$p_T$ hadron production in \cite{r2,r9}, for
 deep-inelastic lepton
scattering in
 \cite{r10} - \cite{kn2}, and for
 hadroproduction of
leading particles on nuclei in
 \cite{r3,r13}.

\section{Radiative energy loss in vacuum}

 A highly virtual quark
 gradually looses energy for radiation of
 gluons
until final hadrons are produced.
 It was a prominent result
of ref.  \cite{n} that
 the density of radiative energy loss $dE/dt$ is
energy and time independent, in analogy to the
 string model. This
is a consequence of
 time ordering of radiation.  The emission of a
gluon which carries a portion $\alpha$ of
 the light-cone momentum of
the quark and
 transverse momentum $k_T$ takes a time
(called radiation time)
\beq
 t_r\approx
\frac{2\nu}{k_T^2}\alpha(1-\alpha)\ ,
\label{2}
\eeq
where $\nu$ is the energy of the
quark.  Eq.  (\ref{2}) follows from the form of the energy
denominator corresponding to such a fluctuation in
 the infinite momentum
frame.

 In the case of hadron production
with energy $z_h\nu$,
energy conservation forbids radiation of
 gluons with energy
higher than $(1-z_h)\nu$.  With
 this restriction the energy loss during
a time
 interval $t$ reads,

 \beq
 \Delta E_{rad}(t)=
\int_{\lambda^2}^{Q^2} dk_T^2
 \int_0^1 d\alpha\ \alpha \nu
\frac{dn_g}{d\alpha dk_T^2}
 \Theta(1-z_h-\alpha)\ \Theta(t-t_r)
\label{5}
 \eeq
 Here the $k_T$ and  $\alpha$ distribution of
the number of emitted gluons
 is,
$dn_g/d\alpha dk_T^2
 = \epsilon /\alpha k_T^2$, where
$\epsilon=4\alpha_s(k_T^2)/3\pi$ is calculated
 perturbatively.

The integration in eq. (\ref{5}) covers both
 soft and hard radiation.
Usually soft
 hadronization is described
in terms of string
 model
\cite{cnn}.  We, however, model it
 by radiation as well.  To get
sensible results, we fix
 the running QCD coupling $\alpha_s(k_T^2) =
\alpha_s(k_0^2)$ at $k_T^2 \leq k_0^2$. The
 parameter $k_0 \approx 0.7\
GeV$ is chosen to
 reproduce the energy loss density corresponding to
 the
string tension, $dE/dt \approx 1\ Gev/fm$.
 This value of
$k_0$  corresponds quite
well to the shortness of the gluon-gluon
 correlation radius
suggested by lattice
 calculation.

Performing the integrations in eq.  (\ref{5}) in the
small--$\alpha$
 approximation we find

 \beqn
 & & \Delta E_{rad}(t)=
{\epsilon\over 2} t (Q^2-\lambda^2) \Theta(t_1-t) +
 \epsilon\nu
(1-z_h)\Theta(t-t_1) + \nonumber\\
 & & \epsilon\nu (1-z_h) \ln\left ({t\over
t_1}\right )
 \Theta(t-t_1) \Theta(t_2-t) +
 \epsilon\nu (1-z_h) \ln\left
({Q^2\over \lambda^2}
 \right ) \Theta(t-t_2),\
 \label{6}
 \eeqn
 where $t_1=(1-z_h)/x_{Bj}m_N$,
 $t_2={Q^2\over
\lambda^2}t_1$, and
$x_{Bj}=Q^2/2m_N\nu$ is the
Bjorken
 variable.

 Up to the time $t=t_1$, the density of energy
 loss
is constant, $dE/dt=-\epsilon Q^2/2$, like
when there is no restriction on
the radiated
 energy \cite{n,kn1,kn2}.
  Then it  slows down to
 $dE/dt=-\epsilon\nu
 (1-z_h)/t$.

Eventually, no radiation is permitted at
$t > t_2$. However,
 a color charge cannot
propagate long
 time without radiation. This will be insured
by the
Sudakov's type formfactor,
 $F(t)=\exp\left[-\tilde
n_g(t)\right]$, where
 $\tilde n_g(t)$ is the number
 of non-radiated
gluons,

 \beq
 \tilde n_g(t)=\epsilon \left[{t\over t_1}-1
 -\ln\left
({t\over t_1}\right )\right ]
 \Theta(t-t_1)
\label{77}
\eeq

 We have arrived at the central question,
at what time is the leading hadron
(or, better, a colorless
ejectile) which does not
radiate energy anymore, produced?  To answer this
question one needs a model for hadronization.
  In the large $N_c$ limit,
 each radiated gluon can be replaced by a $q\bar q$
 pair,
and the whole system can be seen as
made of color dipoles.
 The fastest $q\bar q$ dipole produced is to be projected
over the hadron
wave function, $\Psi_h(\beta,l_T)$,
 where $\beta$ and $1-\beta$ are the
relative
 shares of the light-cone momentum carried by the
 quarks, and $l_T$
is the relative transverse
 momentum of the quarks.  The
 corresponding fragmentation
 function of the quark
into the hadron
can be represented in the
 form,
$ D(z_h) = \int_0^{\infty} dt
 W(t,\nu,z_h)$,
 where $W(t,\nu,z_h)$ is a distribution function of the
 hadron production time
intervals, which reads,
 \beqn
 W(t,\nu,z_h)& \propto &
 \int \limits_0^1
\frac{d\alpha}{\alpha}\
 \delta \left
 [\alpha-2\left
(1-\frac{z_h\nu}{E_q(t)}\right )\right ]
 \int \frac{dk_T^2}{k_T^2}\
\delta\left
 [k_T^2-\frac{2\nu}{t}
 \alpha(1-\alpha)\right ]\times
\nonumber\\
 & & \int dl_t^2\ \delta\left [l_T^2-{9\over
 16}k_T^2 \right ]
\int\limits_0^1 d\beta
 \delta\left [\beta-\frac{\alpha}{2-\alpha}
 \right ]
|\Psi_h(\beta,l_T)|^2
 \label{9}
 \eeqn

 Here
 the quark energy $E_q(t)
= \nu-\Delta E_{rad}(t)$

The $\delta$-functions in eq. (\ref{9}) come from
the conservation of longitudinal and transverse momenta and
from eq. (\ref{2}).

 The hadronic wave function in the
light-cone
 representation, is chosen to satisfy the
 Regge end-point
behaviour $|\Psi_h(l_T^2,\beta)|^2
 \propto \sqrt{\beta} (\sqrt{1-\beta}) (1
+
 l_T^2r_h^2/6)^{-1}$, where $r_h$ is the mean
 square root electromagnetic
radius of the hadron.

 Some examples of the distribution
over time of production,
$W(t,\nu,z_h)$ are
shown in fig.  1.  The mean production time
 $t_{pr}=\int t W(t,\nu,z_h)dt$
approximately scales in
 $(1-z_h)\nu$ and weakly depends on $Q^2$.
It vanishes at $z_h \approx 1$.

Once we know $W(t,\nu,z_h)$ we can calculate
$D(z_h)$, which nicely agrees with the data
from the EMC experiment
 \cite{emc}.

\section{Nuclear medium}

 As soon as the colorless wave packet with the
 desired (detected) momentum
is produced inside the
 nucleus, no subsequent inelastic interaction is
permitted, otherwise the leading
quark degrades its
 energy due to further hadronization.
 Such a restriction means a
nuclear suppression of the production rate.

\begin{figure}[h]
\includegraphics{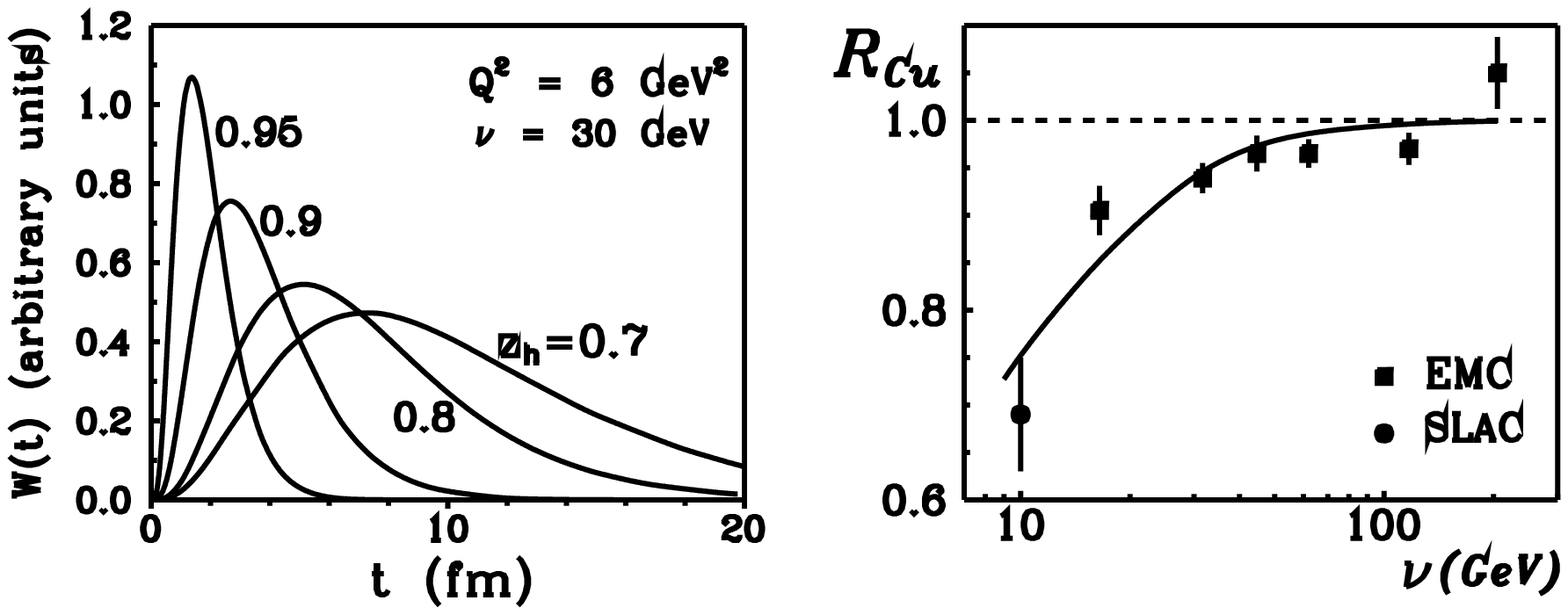}
\begin{center}
\vspace{5.5cm}
\parbox{13cm}
{\caption{Distribution of the production time at $\nu=30\ GeV$
and $Q^2=6\ GeV^2$ vs $z_h=0.7 - 0.95$}
\caption{Comparison of our calculations at $Q^2=6\ GeV^2$ with
the data [14,15] for the $\nu$-dependence
of the ratio of cross sections integrated over $z_h$}}
\end{center}
\end{figure}

\begin{figure}[t]
\includegraphics{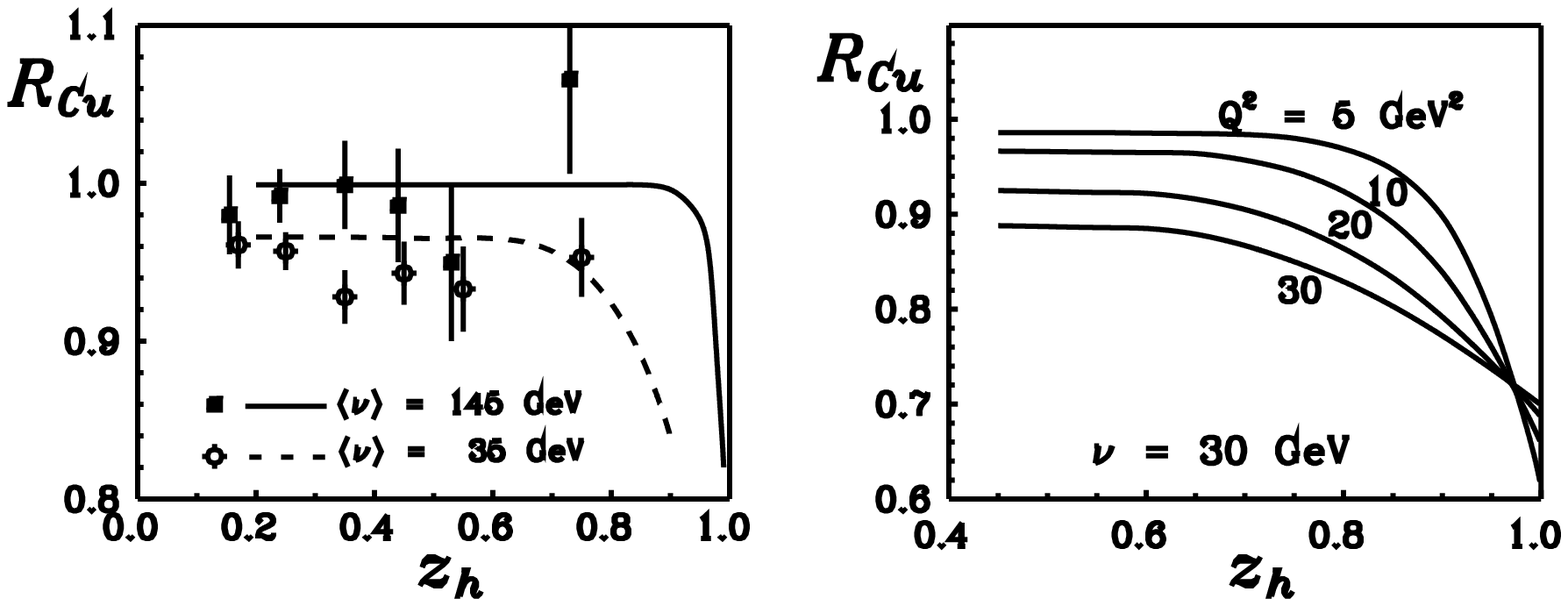}
\begin{center}
\vspace{5.5cm}
\parbox{13cm}
{\caption{$z_h$-dependence of
nuclear transparency. The
data [14] are integrated over $Q^2$}
\caption{Predictions for $z_h$ dependence
of nuclear transparency at $\nu = 30\ GeV$
vs $Q^2$}}
\end{center}
\end{figure}

 On the other hand, soft
interactions
 during hadronization at $t<t_{pr}$ in
 nuclear matter
 cannot
stop or absorb the leading quark \cite{k90}.
Since the quark looses much more energy for
the hard gluon
radiation following the deep-inelastic
scattering
than for the induced soft radiation,
the latter is a
 small
correction, provided only that $Q^2$ is
large.

 The produced colorless wave packet may have a small
 transverse size $\langle \rho^2 \rangle
\approx
 4/\langle l_T^2 \rangle$.  In this case color
 transparency
increases its chances to escape the
 nucleus without interaction.
We also take into accout time evolution
of the produced wave packet.

 The results of our parameter-free calculations are
 compared in fig. 2
with available data \cite{emc,slac} on the energy dependence of
 nuclear transparency integrated
 for $z_h > 0.2$.

The $z_h$-dependence of nuclear transparency is
supposed to be most sensitive to the underlying
 dynamics. We compare our
predictions with the EMC
 data in fig. 3.

 We predict $Q^2$-dependent effects, which
 agree with the
observed weak $Q^2$-dependence of
 nuclear transparency observed by the EMC
experiment \cite{emc}. Much stronger
 effects are expected at ELFE energies, as
shown in fig. 4.

\medskip

In conclusion, we have developed an approach to
 electroproduction of
leading hadrons on nuclei,
based on perturbative QCD.  The results
of parameter-free calculations agree well with
the available data.  We expects
that nuclear effects
 are most sensitive to the underlying dynamics of
hadronization in the energy range of ELFE or
 HERMES

\medskip

{\bf Acknowledgemens}: B.K. is grateful to S.~Bass
for invitation to the Workshop and a partial support. A support from
Max-Planck-Institut f\"ur Kernphysik is also acknowledged.

\end{document}